\documentstyle[11pt,newpasp,twoside,epsfig]{article}
\def\edcomment#1{\iffalse\marginpar{\raggedright\sl#1\/}\else\relax\fi}
\reversemarginpar
\newcommand{\Msolar}{\mbox{\,$\rm M_{\odot}$}}        
\newcommand{\gtsim}{\mbox{{\raisebox{-0.4ex}{$\stackrel{>}{{\scriptstyle\sim}}
$}}}}

\begin{document}
\title{Using MgII to investigate quasars and their black-hole masses}
\author{Matt J.~Jarvis}
\affil{Astrophysics, Department of Physics, University of Oxford, Keble Road, Oxford, OX1
3RH, UK.}
\author{Ross J.~McLure}
\affil{Institute for Astronomy, University of Edinburgh, Royal
Observatory, Edinburgh EH9 3HJ.}

\begin{abstract}

We highlight the importance of the MgII$\lambda 2800$ emission-line
doublet in probing high-redshift quasars and their supermassive black
holes. In the SDSS era, where large scale investigations of quasars
across the age of the Universe are possible, this emission-line has the
ability to provide accurate systemic redshifts which are important for
a variety of follow-up studies, as well as probe the masses of the
supermassive black holes that power these phenomena.

\end{abstract}

\section{Introduction}

The Sloan Digital Sky Survey will provide us with the largest database
of quasars in existence, over all cosmic epochs. Therefore, we now
need to find useful diagnostic tools in order to study these highly
energetic phenomena at all redshifts. 

Various properties of quasars can now be investigated with their
optical spectra alone. As discussed in this meeting, these include orientation, dust 
composition,
the accretion disk, the broad-line region, the narrow-line region and
the black-hole masses. In this contribution we highlight the usefulness of the MgII
emission-line doublet as a diagnostic of black-hole masses in
quasars at $z > 0.3$ and also as a means to determining an accurate
systemic redshift for follow-up studies at other wavelengths.

\section{Estimating black-hole masses in quasars with MgII}

\subsection{The virial estimator}

In recent years estimating black-hole masses has become somewhat
easier than in the past. This is due to the large amount of work
carried out on reverberation mapping of Seyfert galaxies and quasars
(Wandel, Peterson \& Malkan 1999; Kaspi et al. 2000; and see Peterson these proceedings) and the
correlation discovered by Kaspi et al. between the radius of the
broad-line region and the monochromatic luminosity at
5100\AA. If the velocity of the broad-line region gas can
be measured along with the luminosity then the mass of the dominant
gravitational mass can be estimated, assuming that the motion is
virialised via, $M_{bh}=G^{-1}R_{BLR}V_{BLR}^{2}$,
where the velocity of the broad line region is usually derived from the FWHM of the broad-emission lines, traditionally H$\beta$.

However, using this technique to estimate the masses of black holes in quasars at high 
redshift ($z > 0.9$) is difficult because
H$\beta$ is redshifted out of the optical passbands and into the
near-infrared where observations are much more difficult. Therefore, to probe the 
high-redshift regime we need a proxy for H$\beta$ and $\lambda L_{5100}$ in the ultraviolet region of the spectrum. An obvious choice for this proxy is MgII$\lambda 2800$.

\subsection{MgII as a proxy for H$\beta$}

MgII$\lambda 2800$ has a similar ionisation potential to that of
H$\beta$ and thus we would expect that the line emission arises from a
similar distance from the central ionising source. MgII is far enough in the ultraviolet 
part of the spectrum that it can be seen in optical spectra up to redshift $z \sim 2.3$. 
Thus, calibrating the use of this line with H$\beta$ allows us to estimate black-hole masses over the majority of the age of the Universe.

A detailed account of calibrating MgII with H$\beta$ to estimate
black-hole masses in quasars is given in McLure \& Jarvis (2002). We
will only summarize this work here. The correlation between the radius
of the broad-line region, $R_{\rm BLR}$, and the monochromatic
luminosity at 3000\AA\, is given by,

\begin{equation}
R_{\rm BLR}=(26.1\pm3.6) \left[\lambda L_{3000}/10^{37}W
\right]^{(0.50\pm0.02)}.
\end{equation}
This leads to\footnote{This relation has now been updated for
 genuinely powerful, high-redshift quasars only (see McLure \& Dunlop 2003)},

\begin{equation}
\frac{M_{\rm bh}}{\Msolar}  =3.21\left(\frac{\lambda
L_{3000}}{10^{37}{\rm W}}\right)^{0.5}\left(\frac{\rm FWHM(MgII)}
{{\rm kms}^{-1}}\right)^{2}.
\label{final}
\end{equation}

By compiling spectra covering both H$\beta$ and MgII of the
reverberation mapped samples of Wandel et al. (1999) and Kaspi et
al. (2000) we are able to directly compare the UV virial black-hole mass
estimate (using MgII and monochromatic luminosity at 3000\AA) and the
traditional method using H$\beta$. Fig.~1a shows that the FWHM(MgII)
closely follows the FWHM(H$\beta$) in these spectra, as expected
if the lines are emitted from the same region. 

\begin{figure}[!h]
\plotone{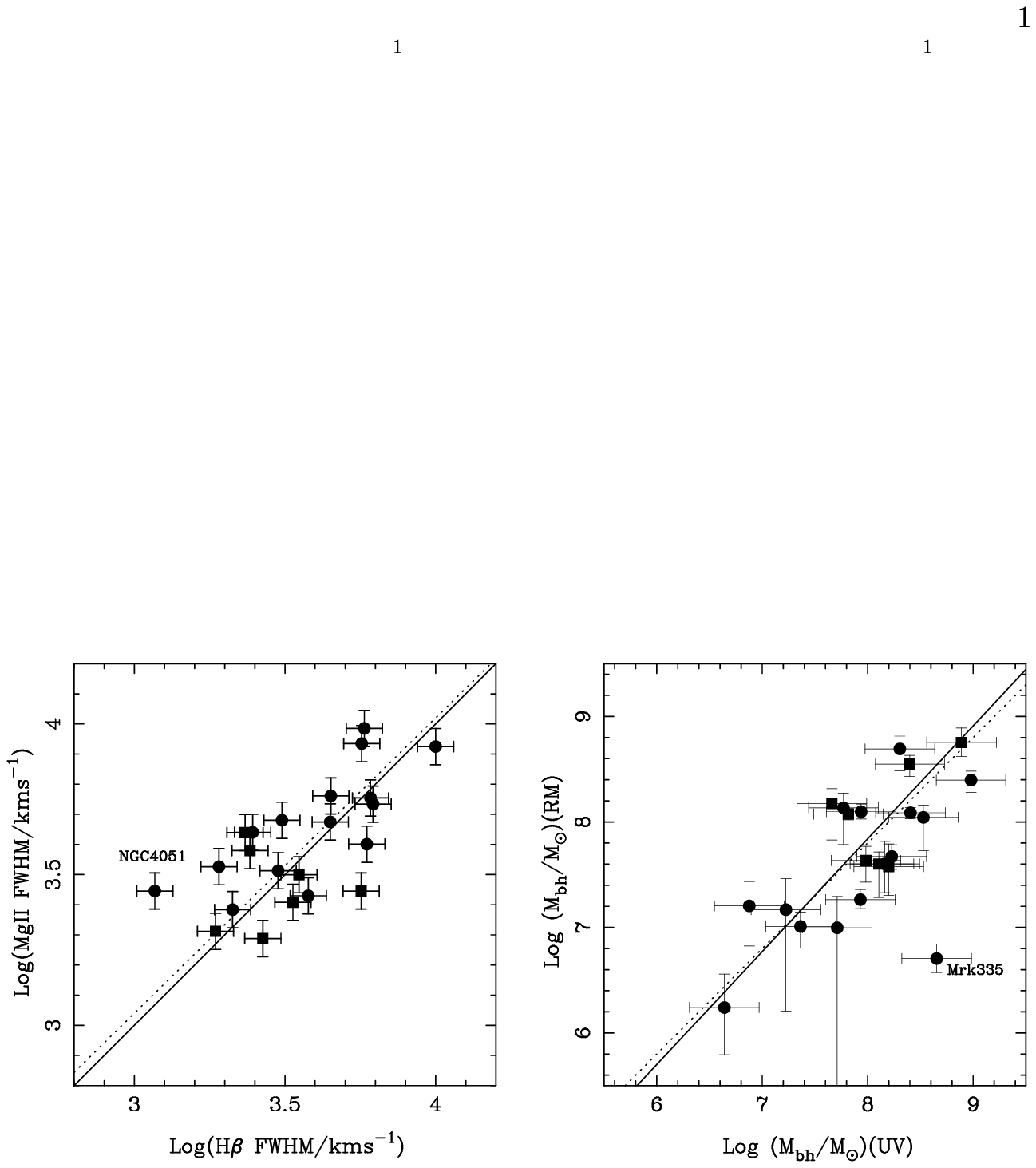}
\caption{({\it left}) FWHM(H$\beta$) versus FWHM(MgII) for a sub-set of the 
  reverberation mapped (RM) samples of Wandel et al. (2000) and Kaspi et
  al. (2000). ({\it right}) Black-hole masses measured directly from reverberation 
mapping versus the black-hole mass estimate using the virial technique with MgII and $\lambda L_{3000}$ from McLure \& Jarvis (2002).}
\end{figure}

\section{The dependence of radio power on black-hole mass}

We can now use the virial estimator to calculate $M_{\rm bh}$ in quasars over all cosmic epochs using optical or
near-infrared spectroscopy alone. This gives us sufficient numbers of
objects to probe the black-hole masses in different populations of
quasars, i.e. the radio-loud quasars and the radio-quiet quasars. In
Fig.~2a we plot the black-hole masses derived from spectra of the Large
Bright Quasar Survey [LBQS; Hewett, Foltz \& Chaffee (1995)], predominantly
comprised of radio-quiet quasars  and the Molonglo quasar
sample (Kapahi et al. 1998), which is a low-frequency radio selected sample. One can
easily see how the radio-loud quasars almost exclusively lie to the
top right of the plot, whereas the radio-quiet quasars from the LBQS
stretch out across the full range of black-hole mass.

Therefore, there does seem to be evidence for the genuinely
powerful ($L_{5 \rm GHz} > 10^{24}$~W~Hz$^{-1}$~sr$^{-1}$) radio-loud
quasars to have black-holes confined to upper mass range of $M_{\rm
  bh}\, \gtsim\, 10^{8}\Msolar$, in agreement with Dunlop et al. (2003). Flat-spectrum radio-loud quasars do not necessarily conform to this
suggestion (e.g. Oshlack, Webster \& Whiting 2002), but this may be
explained by the consideration of source geometry and Doppler boosting
of the radio-flux in these sources (Jarvis \& McLure 2002; Fig.2b).

\begin{figure}[!h]
\plotone{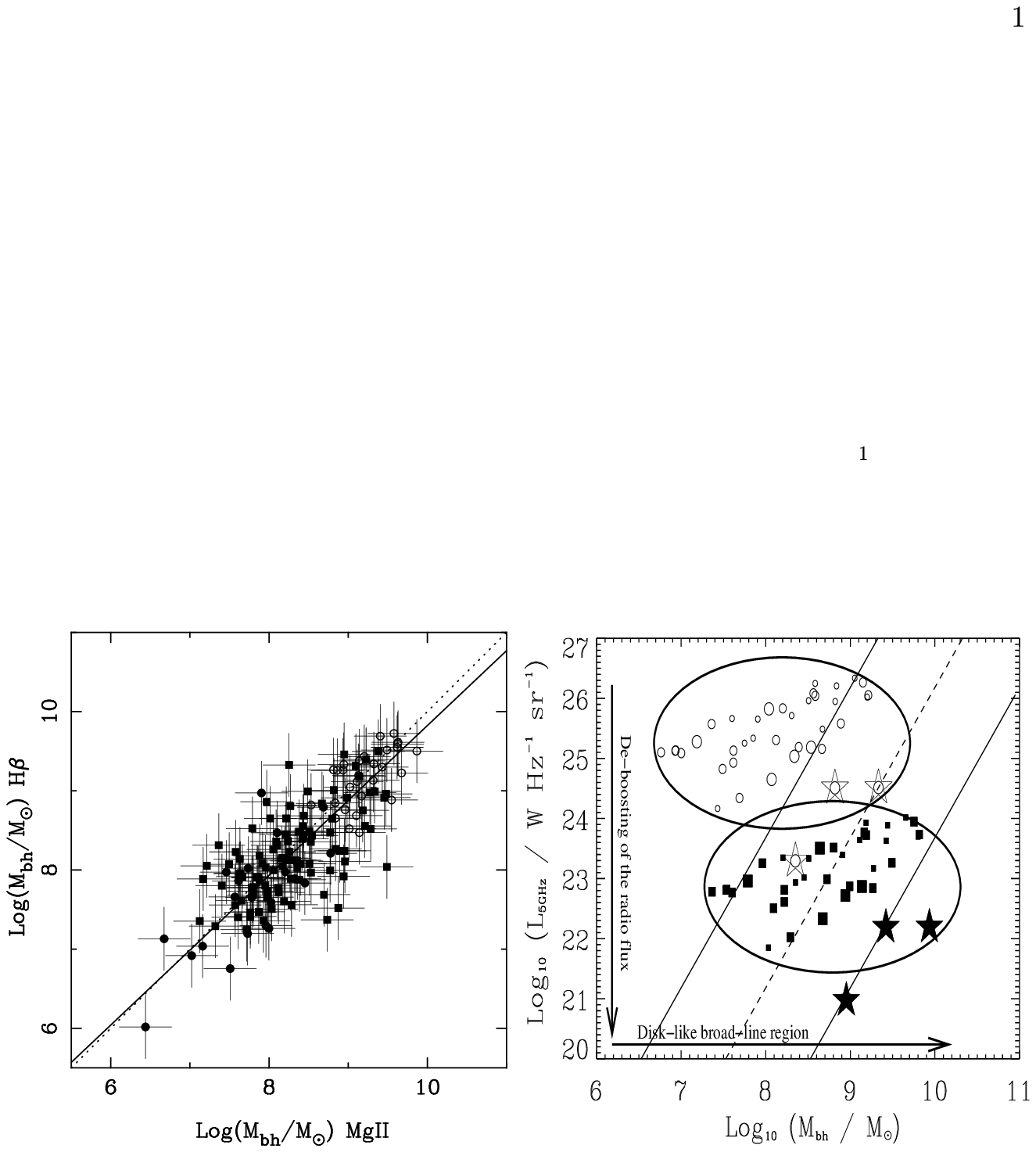}
\caption{ ({\it Left}) The H$\beta$ versus Mg{\sc ii} virial black-hole
estimators for 150 objects from the RM (filled circles), LBQS (filled
squares) and MQS (open circles) samples. The dotted line is an 
exact 1:1 relation. The solid line is the
BCES bisector fit which has a slope of $0.95\pm0.06$ (McLure \& Jarvis
2002). ({\it Right}) The $M_{\rm bh}$ -- $L_{\rm rad}$ plane
adapted from Jarvis \& McLure (2002). Open symbols are
from the study of Oshlack et al. (2002), the filled symbols represent
where these points would lie after corrections for Doppler boosting
and a disk-like geometry for the
broad-line region. The large stars are anomalous steep-spectrum
objects and are probably not dominated by the core emission. }
\end{figure}

\section{Accurate redshifts using MgII}
In addition to using MgII as a black-hole mass estimator in quasars,
it can also be used to provide accurate quasar redshifts, which
are important in various follow-up studies of the quasar host
galaxies and the neutral gas within the epoch of reionization.

Recently Walter et al. (2003) have detected the presence of molecular
gas in the quasar SDSS J1148+5251 at $z \sim 6.42$. This kind of
search is only efficient when relatively accurate redshifts are known
due to the limited bandwidth of molecular line searches in the radio
regime. 

Accurate redshifts are also required to determine the fraction of
neutral hydrogen to the foreground of the high-redshift $z > 6$
quasars which exhibit the Gunn-Peterson trough (Becker et al. 2001). This is extremely difficult with Lyman-$\alpha$
alone due to the significant absorption blueward of the Lyman-$\alpha$
emission line. The accurate redshifts derived from MgII are
relatively easy to obtain and thus, may be very important in future
studies.



\references

Becker, R.L., et al., 2001, \aj, 122, 2850\\
Dunlop, J.S., et al., 2003, \mnras, 340, 1095 \\
Hewett, P.C., Foltz, C.B., Chaffee, F.H., 1995, \aj,
109, 1498\\
Jarvis, M.J., \& McLure, R.J., 2002, \mnras, 336, L38 \\
Kapahi, V.K., et al., 1998, \apjs, 118, 327\\
Kaspi, S., et al., 2000, \apj, 533, 631 \\
McLure, R.J., \& Jarvis, M.J., 2002, \mnras, 337, 109  \\
McLure, R.J., \& Dunlop, J.S., 2003, \mnras, submitted (astro-ph/0310267)\\
Oshlack A., Webster R., Whiting M., 2002, \apj, 576, 81O\\
Walter, F., et al., 2003, Nature, 424, 406\\
Wandel, A., Peterson B.M., Malkan M.A., 1999, ApJ, 526, 57\\
\end{document}